\documentclass[a4paper,11pt]{article}
\usepackage{jinstpub} 
\usepackage{lineno}

\title{\boldmath The UV Laser Calibration System for measuring the Electric field in the SBND}

\author{Shivaraj Mulleria Babu On Behalf of SBND Collaboration}
\affiliation{University of Bern,\\
Sidlerstrasse 5 \\
3012 Bern, Switzerland}
\affiliation{}

\emailAdd{shivaraj.mulleriababu@unibe.ch}

\abstract{ The Short-Baseline Near Detector (SBND) is a LArTPC detector located 110 meters from the Fermilab’s Booster Neutrino Beam (BNB). It is designed to measure neutrino cross sections and aid in excess electron-like neutrino searches. The electric field inside the SBND-TPC may have distortions due to a number of reasons, such as the space charge effect. The space charge effect comes from the abundance of cosmic rays that ionize the argon, producing copious positive argon ions. In this case, we need an alternative solution to determine the electric field distortion inside the TPC volume and compensate for the possible distortion in the spatial information. The UV calibration system is one such attempt to determine the electric field distortion. By utilizing a high-energy ultraviolet laser beam, the system can map the spatial distortions within the TPC volume and provide the necessary corrections to ensure accurate spatial and calorimetric measurements. }


\arxivnumber{} 

\begin{document}
\maketitle
\flushbottom

\section{Introduction}
\label{sec:intro}

\quad The Short-Baseline Near Detector (SBND) is a Liquid Argon Time Projection Chamber (LArTPC) built to be the near detector for the Short-Baseline Neutrino (SBN) program at Fermilab \cite{sbnd}. Located just 110m from the Booster Neutrino Beam (BNB) target, SBND is uniquely positioned to record a large number of neutrino interactions each year, providing more data than any previous LArTPC experiment. This data will help us better understand neutrino interactions with argon and explore new physics beyond the Standard Model with it's high statistical data.

\quad The SBND is designed to operate two Time Projection Chamber (TPC), two drift volumes separated by a central cathode with 110 tons of liquid argon in its active volume. Each TPC has a dimension of 2 m $\times$ 2 m $\times$ 5 m. This design enables detailed 3D imaging of particle interactions, allowing us to track particles with incredible precision and measure their energy. In July 2024, the TPC reached a major milestone when it was powered up to its nominal voltage of $-$100 kV. Working alongside the TPC is the Photon Detection System (PDS), which uses 120 photomultiplier tubes (PMTs) and 192 X-ARAPUCAs to detect scintillation light from particle interactions. These light signals are critical for providing nanosecond-level timing for triggering and event reconstruction. The Cosmic Ray Tagger (CRT) surrounds the detector with scintillator panels, enabling precise tagging of cosmic rays to separate background noise from neutrino signals. 

\quad A key challenge for SBND is mitigating distortions in the electric field within the TPC, primarily caused by the space charge effect(SCE) by the accumulation of positive argon ions generated by cosmic ray activity. These distortions impact the spatial and calorimetric reconstruction of particle interactions, essential for achieving SBND’s scientific goals. To address this, SBND utilizes a UV Laser Calibration System(LCS), which maps electric field distortions using ultraviolet laser beams to create ionization tracks. This proceeding discuss the design, development, hardware and electrical details of the system focusing on the recent milestones.



\section{UV Laser Calibration System Overview}

\quad The UV laser calibration system for the SBND utilises a finely tuned, high-energy Class-4 Nd:YAG laser from Continnum-Surelite to create controlled ionization tracks within the TPC. The laser operates at a wavelength of 266 nm, generated by shifting infrared light (1064 nm) to green (532 nm) and finally to UV through harmonic generation. Each laser pulse, with an energy of 60 mJ and a repetition rate of up to 10 Hz, ionizes argon atoms along its path, producing well-defined tracks with minimal scattering and no delta-ray emission. These ionization tracks are reconstructed and compared to their expected paths (true track) to measure electric field distortions inside the detector.

\quad Liquid argon has an ionization potential of 13.84 eV \cite{ionisation}, which corresponds to a photon wavelength of approximately 89 nm. However, lasers capable of producing light at such short wavelengths are not commercially available. Instead, ionization is achieved through multiphoton absorption \cite{multiphoton}. For instance, Nd:YAG lasers with harmonic generation at 266 nm, are capable of producing photons with energy $E_\text{photon} \approx$ 4.66 eV, requiring three photons (\( n = 3 \)) to ionize liquid argon.

\section{Design and Strategy}
\begin{figure}[htbp]
\centering
\includegraphics[width=.5\textwidth]{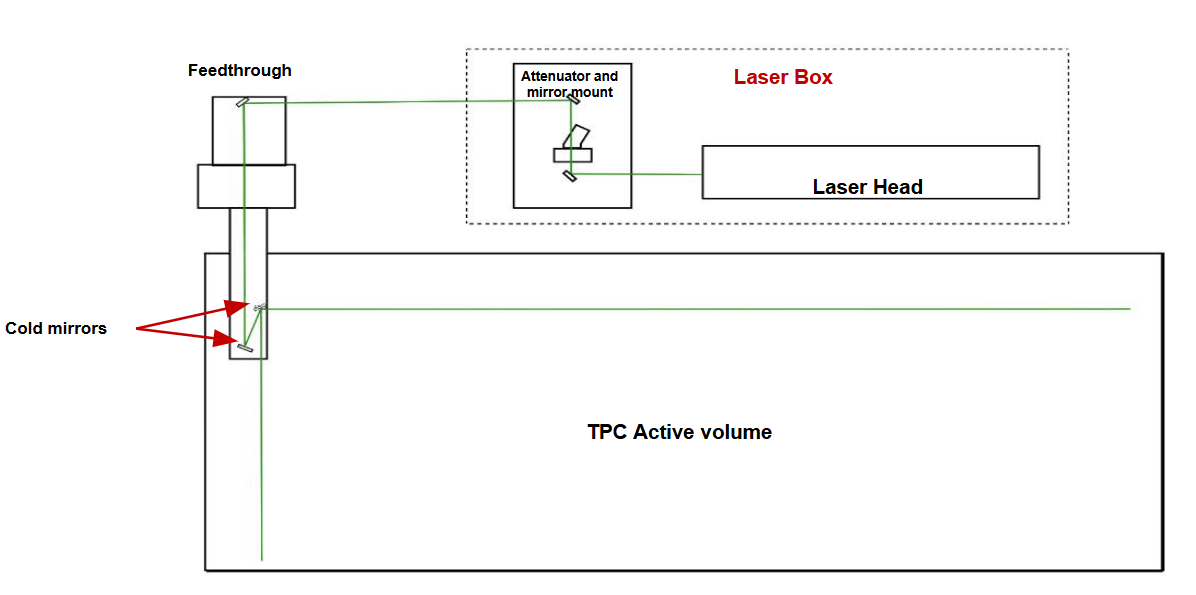}
\qquad
\includegraphics[width=.4\textwidth]{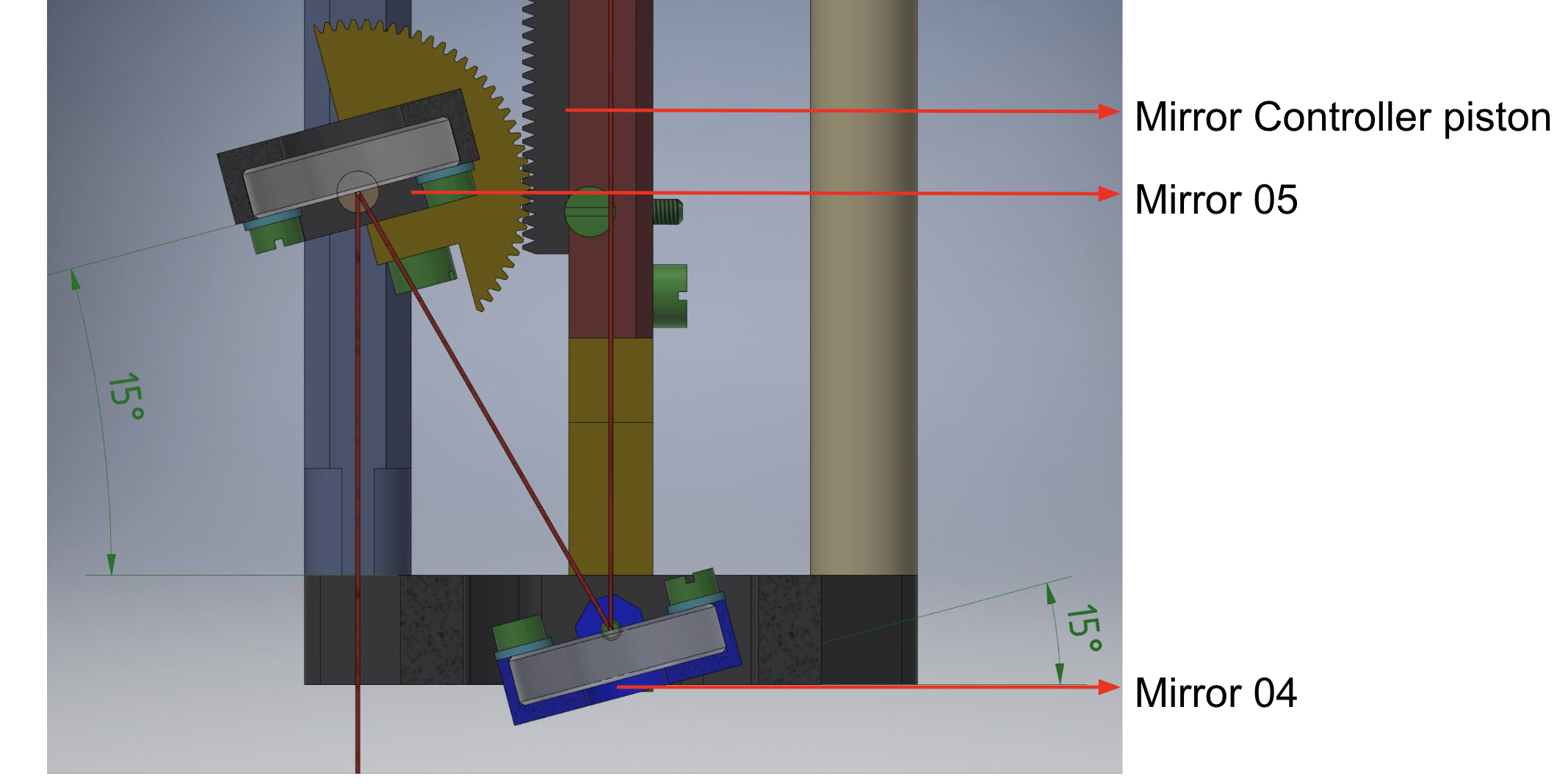}
\caption{(a) Open view Schematics of the laser system at SBND. (b) Close-up view of cold mirrors situated inside the TPC volume.\label{fig:schema}}
\end{figure}

\quad A laser beam is emitted from the laser head and subsequently passes through an attenuator and a series of dichroic mirrors as shown in figure \ref{fig:schema}(a). These mirrors reflect 99\% of 266 nm UV light while transmitting photons of other wavelengths. This process isolates a clean UV beam, which is then directed through the center of a feedthrough into the TPC. 

\quad The feedthrough is mounted with a rotary motor and a high-resolution encoder with a resolution of $0.001^\circ$, enables precise angular control of the laser path inside the LArTPC, while a linear motor with a 3 µm resolution adjusts the vertical position of cold mirrors to accurately guide the laser beam. A 2 m long vacuum-sealed quartz tube passing through the center of feedthrough prevents condensation and ensures clean laser penetration through the argon interface. Using durable materials like polyamide-imide (PAI) the feedthrough maintains stability under extreme low temperatures of liquid argon.

\begin{figure}[htbp]
\centering
\includegraphics[width=.4\textwidth]{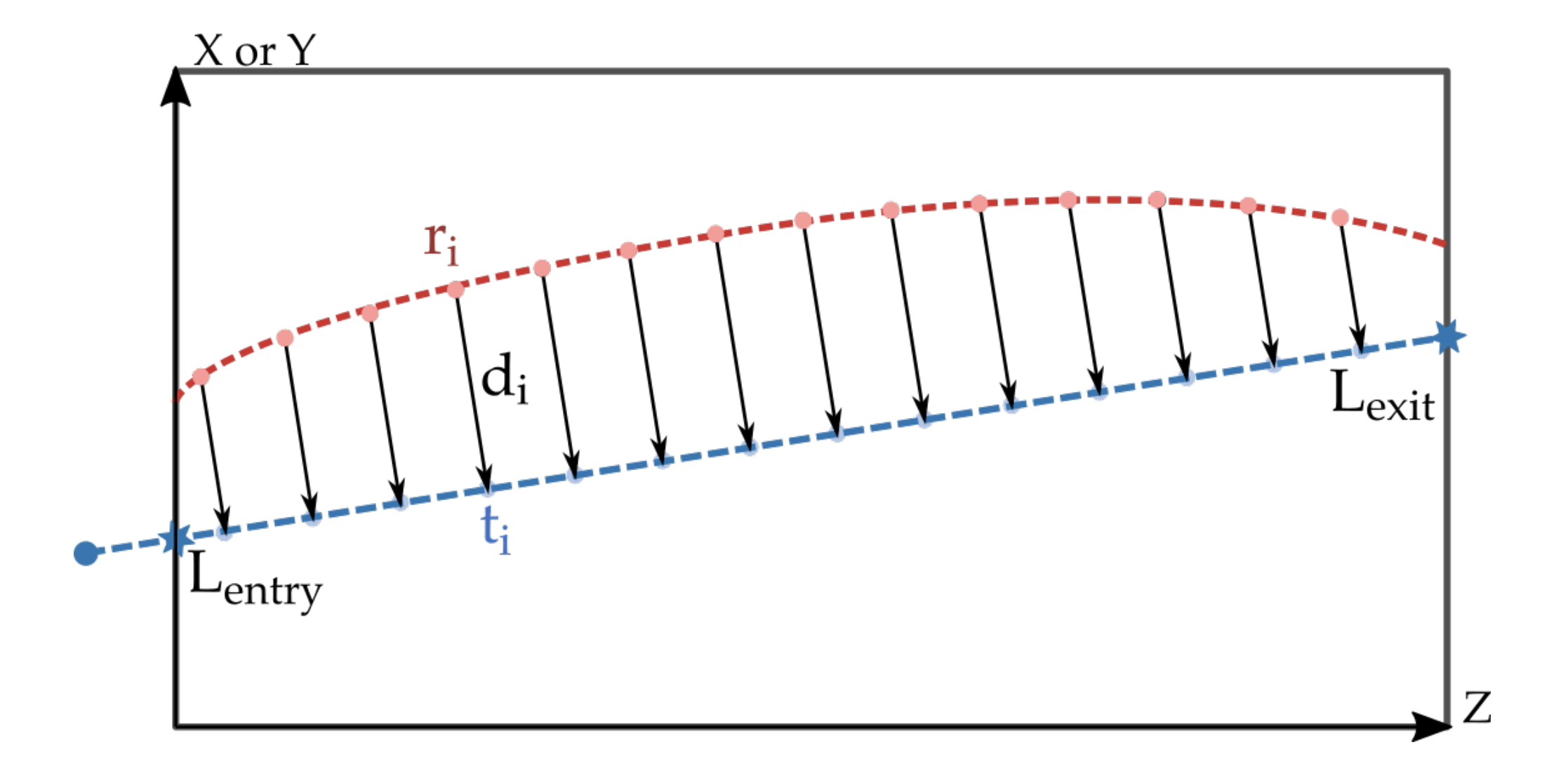}
\qquad
\includegraphics[width=.4\textwidth]{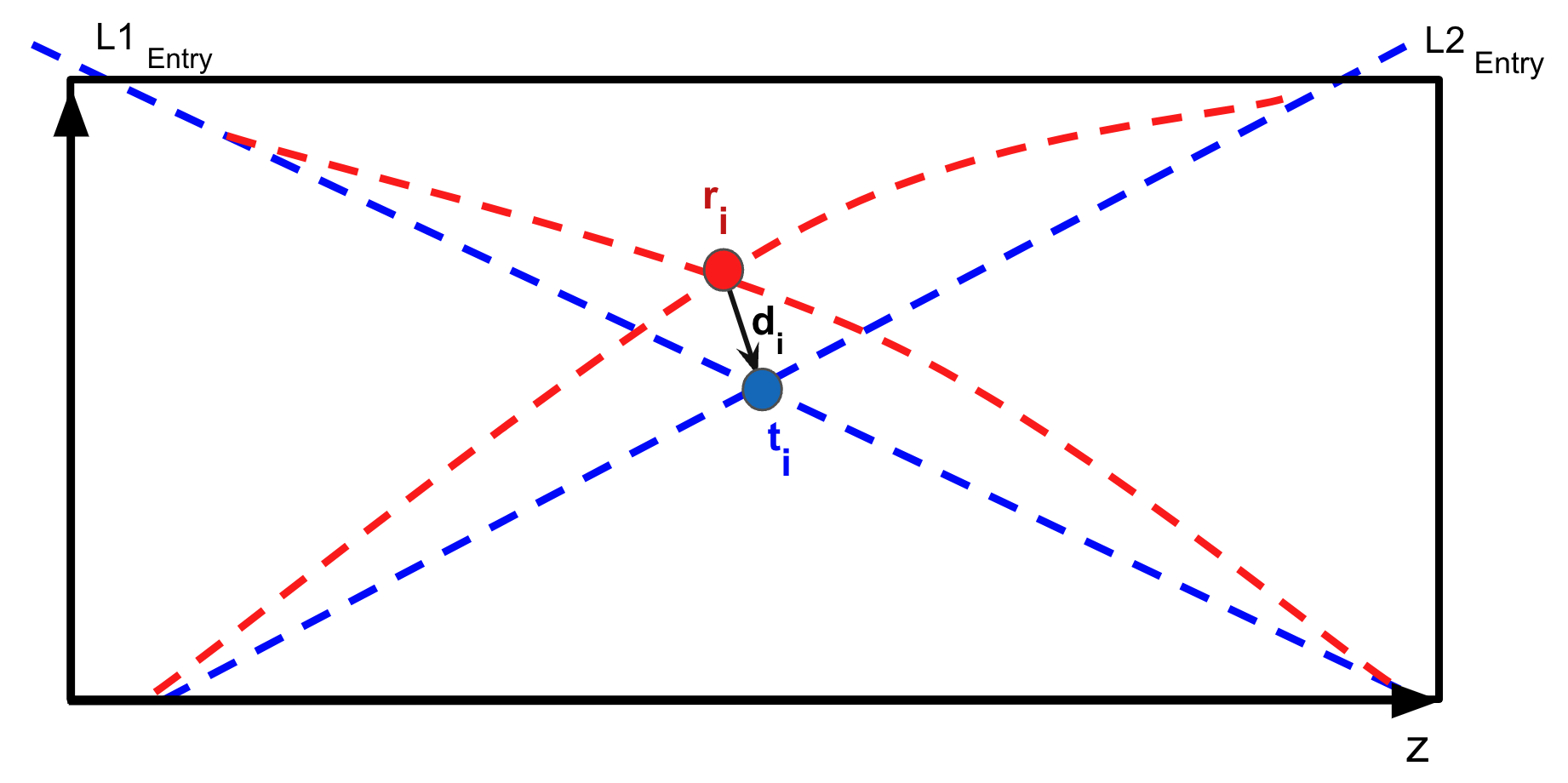}
\caption{(a) Closest point projection method used in MicroBooNE, where blue is the true track info and red is reconstructed track. The displacement vector $d_i$ starting from $r_i$ to $t_i$ is the correction vector. (b) Crossing track approach for SBND calibration system.\label{fig:cross}}
\end{figure}

\quad In MicroBooNE, the closest-point projection method estimates spatial displacement vectors by projecting reconstructed track points perpendicularly onto the true track in 3D space. While effective, this method introduces laser angle dependencies as it forces the displacement vectors to be perpendicular to the true laser tracks, leading to biases that require iterative corrections \cite{uboone_laser}. In SBND, by employing a crossing track method instead of closest-point as illustrated in the figure \ref{fig:cross}(b) the bias is reduced. This approach uses multiple laser tracks that intersect within the TPC, creating unique crossing points that can be easily flagged and used as reference points for comparison. The crossing track points are less susceptible to angle dependencies and offer a clear way to correlate reconstructed points with true track locations. The placement of cold mirrors inside the field cage rings enables crossing points throughout the TPC, further improving spatial resolution.
\section{Electrical Components and Communication}

\begin{figure}[htbp]
\centering
\includegraphics[width=.485\textwidth]{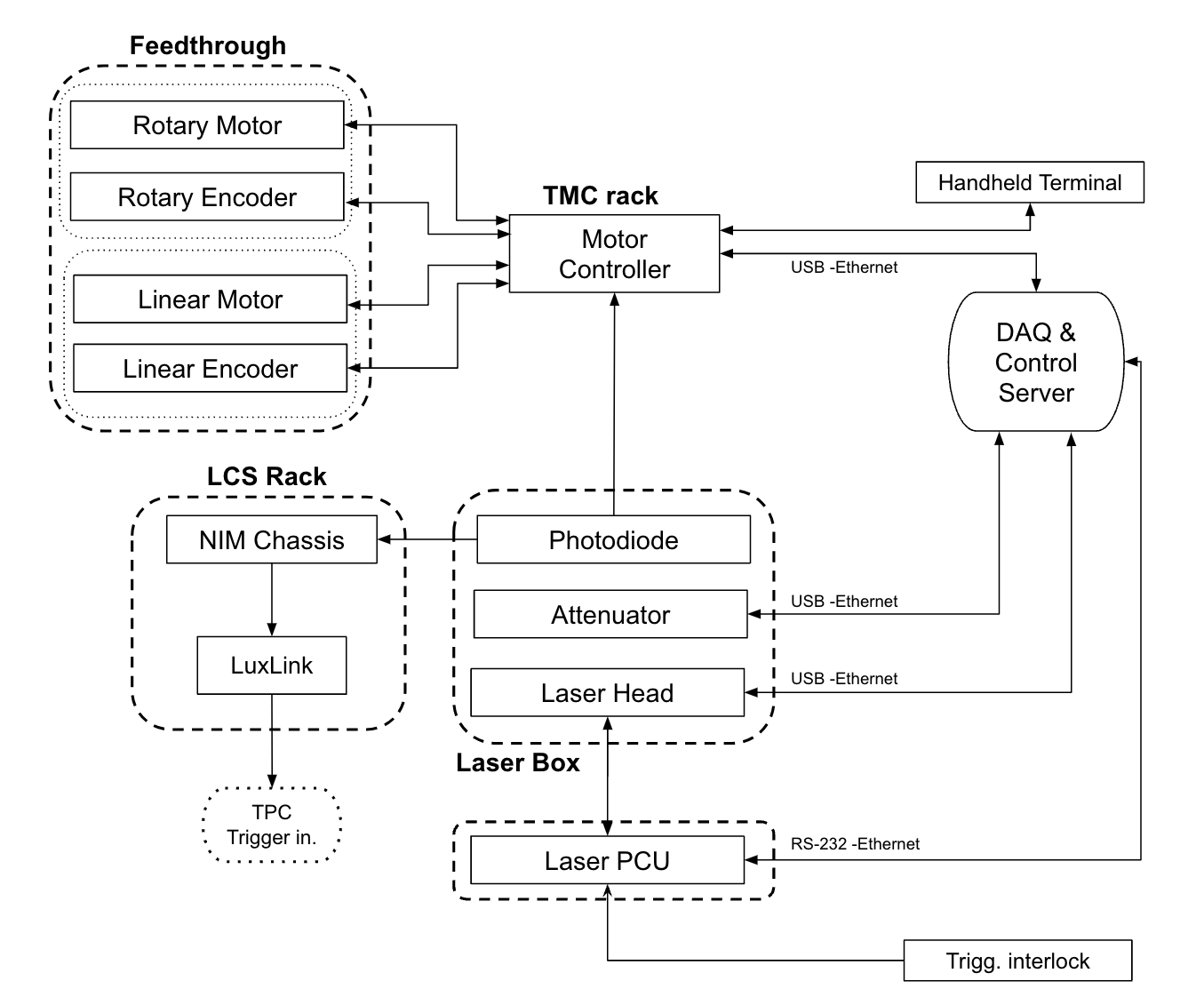}
\qquad
\includegraphics[width=.45\textwidth]{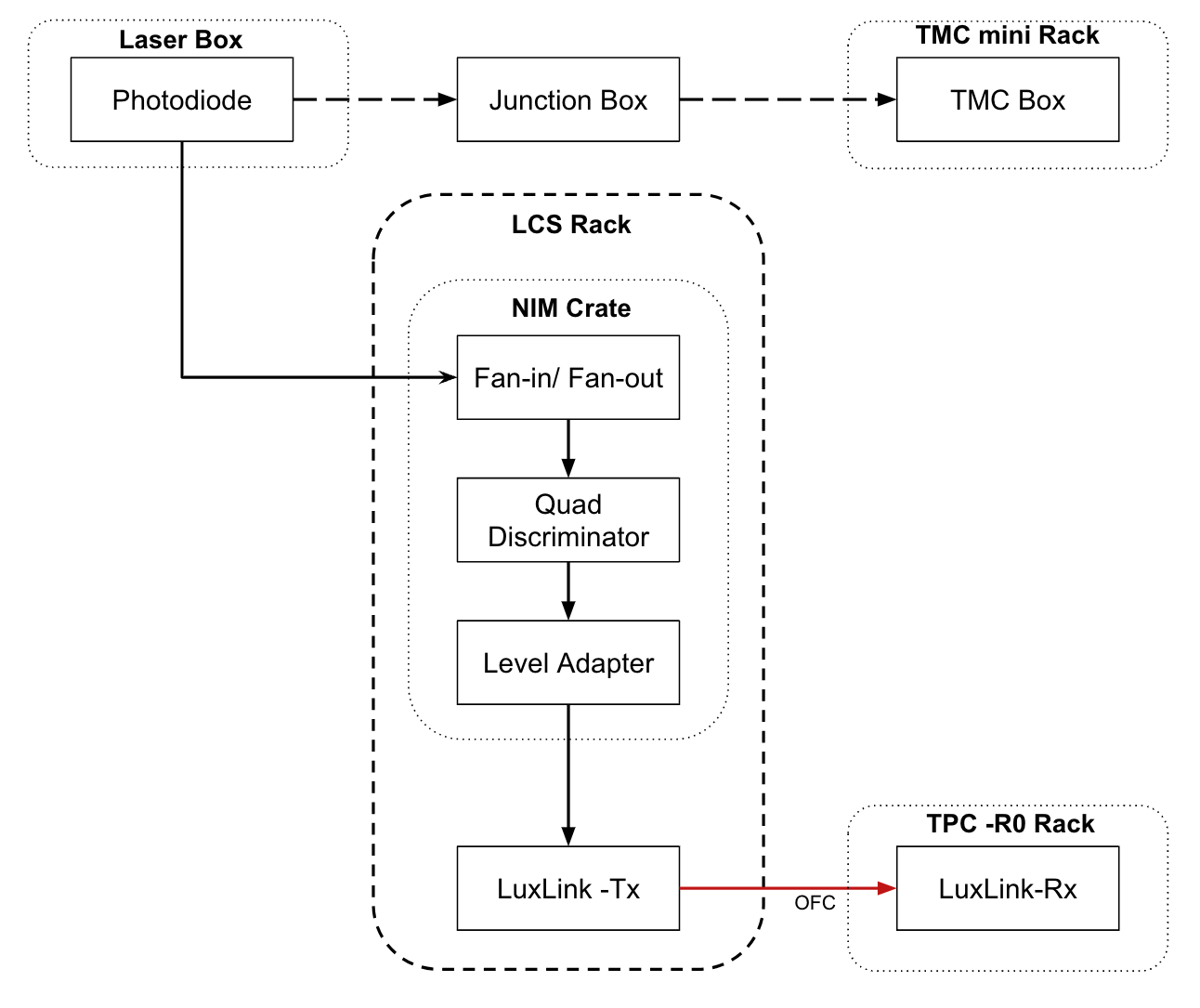}
\caption{(a) Electrical components and connection schematics. (b) Trigger signal processing flow chart.\label{fig:electricals}}
\end{figure}

\quad The electrical system comprises AC and DC power supplies for various components, including a 208 $V_{\text{AC}}$ supply for the laser head and Thermionic Motor Controller (TMC) units. The TMC boxes drive the rotary and linear motors, operating within a voltage range of 24 – 48 $V_{\text{DC}}$, enabling high-resolution beam steering with encoder feedback at µm precision. Low-voltage power is provided by the LV chassis, delivering 5 $V_{\text{DC}}$ and 12 $V_{\text{DC}}$ to auxiliary components such as reference lasers and photodiodes, with inline fuses for overcurrent protection. A photodiode-based triggering system ensures synchronization between the laser and the DAQ by generating a 5 $V$, 50 µs signal upon laser firing, which is processed and transmitted as a 120 ns TTL signal to the DAQ via fiber optic cables, mitigating cross-grounding issues.The Electrical components schematics and the trigger infrastructure is shown in figure \ref{fig:electricals}

\quad The control and communication of the UV laser system are managed through three primary components: the Laser PCU, the TMC Box, and the Altechna Attenuator Controller. These components are connected via Ethernet cables to a central computer server located in the CRT Rack at the SBND building. MCode/TCP with Python is used to facilitate precise control of these components, enabling tasks such as mirror movements, laser parameter adjustments, laser operation and status monitoring. Additionally, a user-friendly interface simplifies operations, providing an intuitive way for users to interact with the system effectively.

\section{Current Status}
\begin{figure}[htbp]
\centering
\includegraphics[width=.4\textwidth]{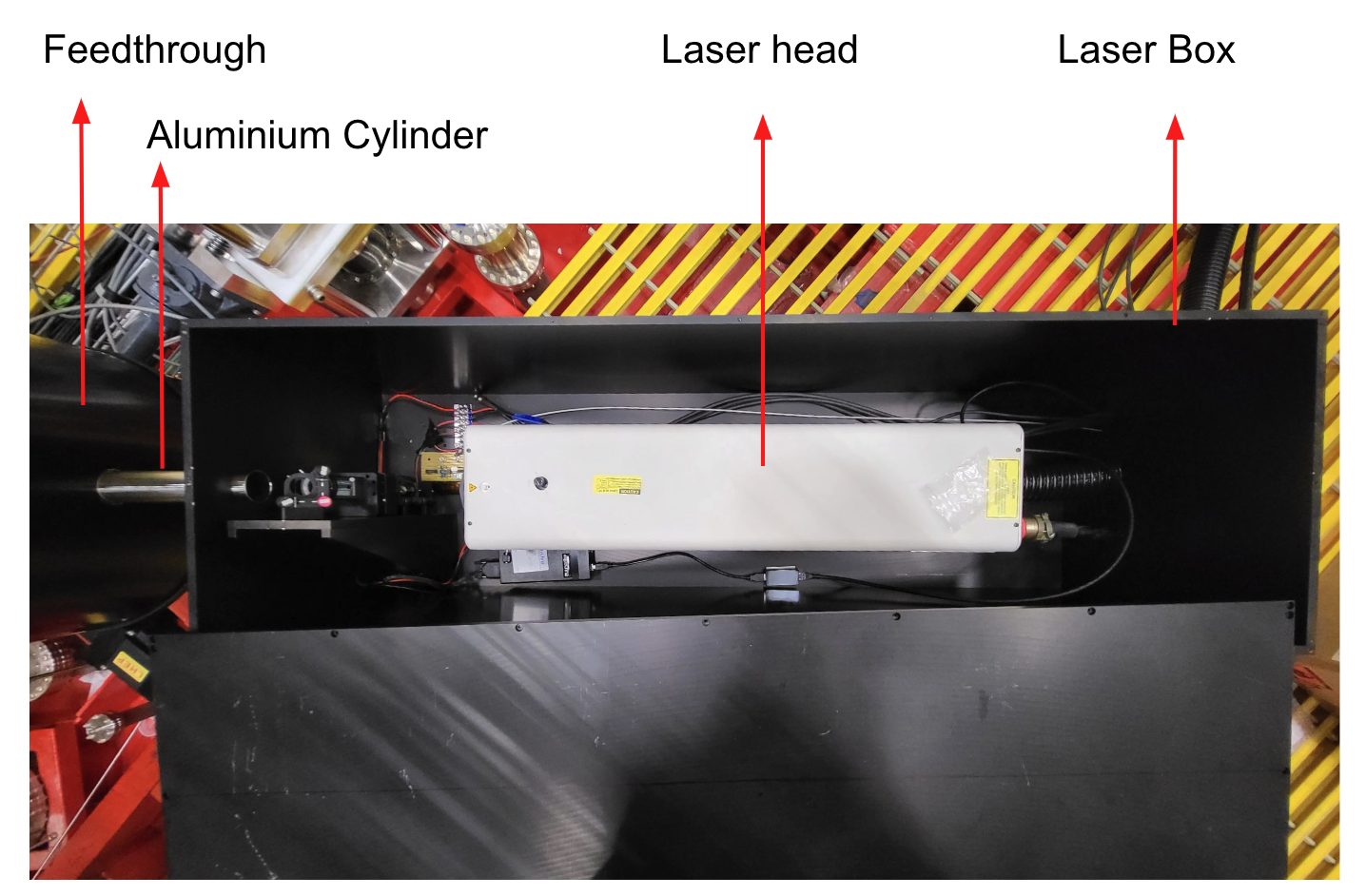}
\qquad
\includegraphics[width=.4\textwidth]{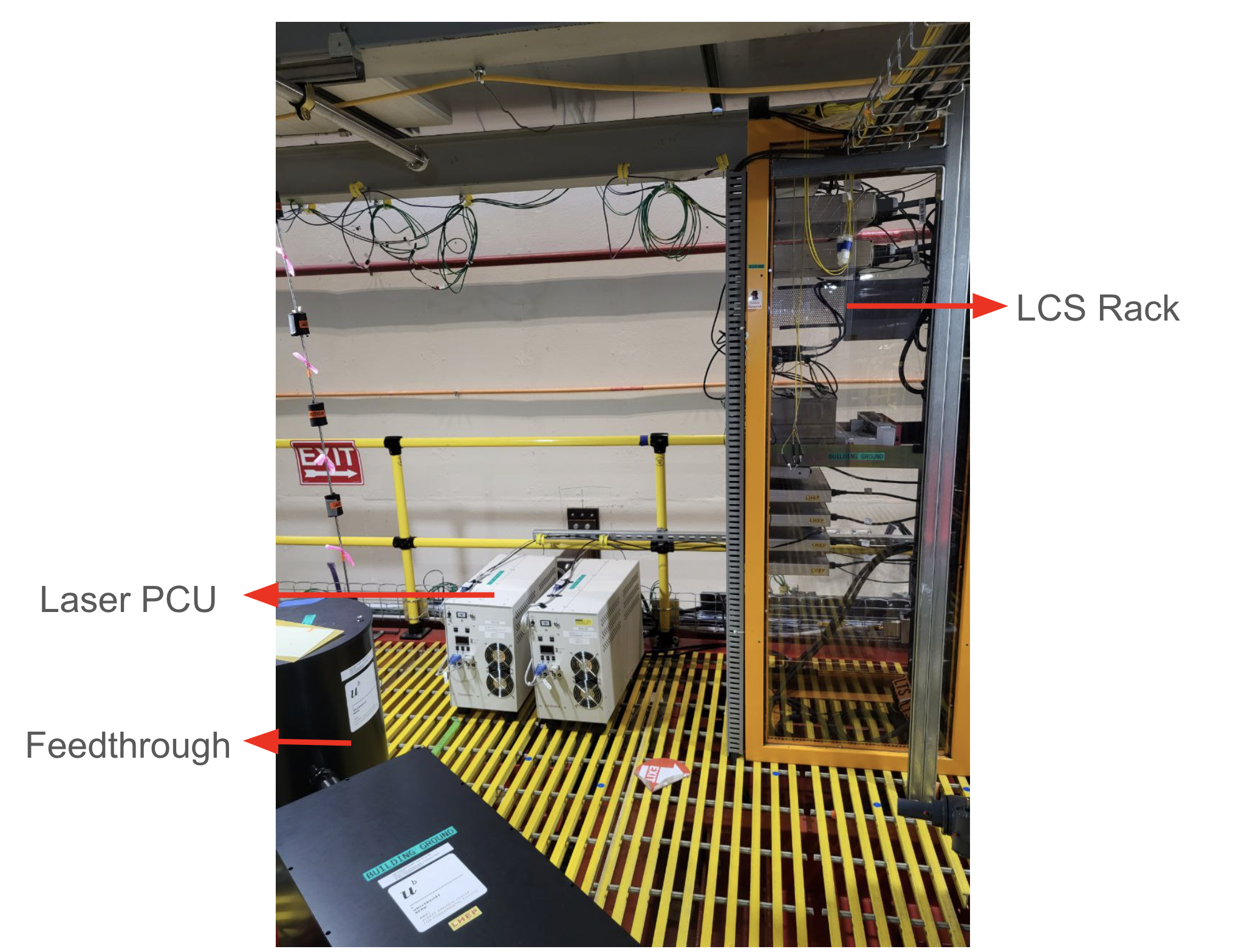}
\caption{(a) Open view of Laser Box (b) Laser system, power supply unit and laser rack on top of SBND\label{fig:install}}
\end{figure}

\quad The SBND UV Laser System has made significant progress toward installation and operational readiness. The entire system was designed, assembled, and tested at the Laboratory for High Energy Physics (LHEP) at the University of Bern in Switzerland before being transported to the Short-Baseline Near Detector (SBND) at Fermilab, where it was installed on the cryostat. The laser system and related equipments has been successfully installed. Images of the installation and the system are shown in Figure \ref{fig:install}. Power and communication cabling have been completed with routes carefully planned, labeled, and organized to ensure safety and efficiency.  The system has successfully undergone operational safety reviews, with approval from electrical, mechanical, and fire safety authorities in September 2024. With these milestones complete, the system is ready for its first laser operations in 2025.

\begin{figure}[htbp]
\centering
\includegraphics[width=.6\textwidth]{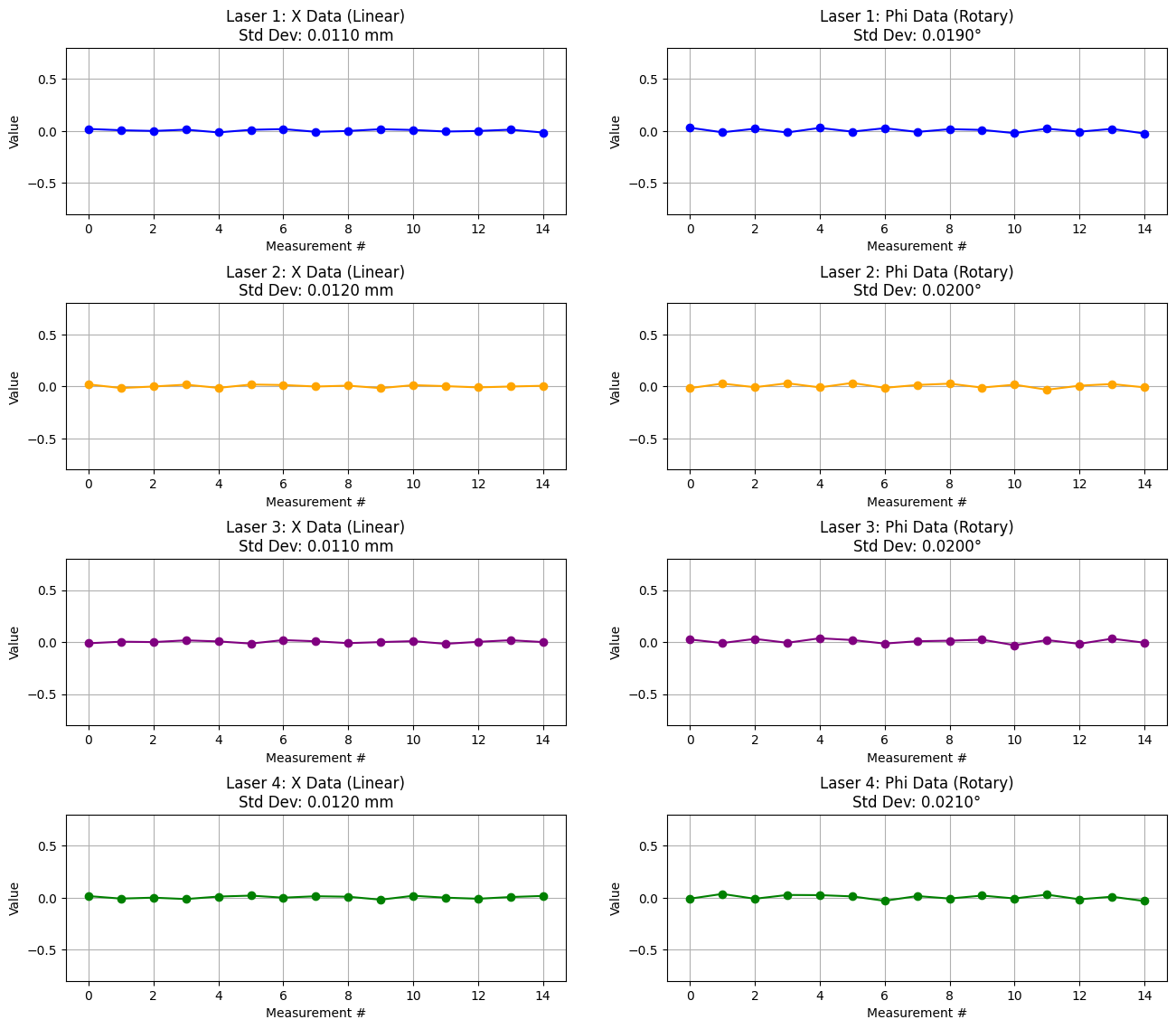}
\caption{Results from repeatability test with the standard deviation for each system.\label{fig:repeat}}
\end{figure}

\quad The system was also tested for repeatability of the mirror position by doing random homing procedure, where the mirrors were moved to random positions and instructed to move to a set home position. The figure \ref{fig:repeat} shows the repeatability measurements for the displacement by linear motor ($x$ in mm) and rotary motor (angular deviation- $\phi$). The results indicate consistent performance, with standard deviations for $x$ ranging from 0.011 mm to 0.012 mm and for $\phi$ between 0.019° and 0.021°. For the rotary motor the spread is higher, which is due to the higher gear ratio and 0.21° spread results in only 2 mm deviation over a distance of 5 m and these values highlights the precision and stability of the system.

\section{Summary}
\quad The SBND UV Laser Calibration System has achieved significant milestones, including successful installation, operational safety approval, and readiness for commissioning in 2025. The system demonstrates precise functionality, with repeatability tests confirming standard deviations of 0.011 - 0.012 mm for linear displacement and 0.019°- 0.021° for angular deviation. These results highlight the system's capability to map electric field distortions with high accuracy, ensuring reliable spatial and calorimetric measurements essential for achieving SBND's scientific goals. Moving forward, the system is well-positioned to enhance detector calibration and contribute to the success of the Short-Baseline Neutrino program.




\providecommand{\href}[2]{#2}\begingroup\raggedright\endgroup







\end{document}